# Band modification in (Ga, Mn)As evidenced by new measurement scheme --- magnetic photoresistance circular dichroism

XueJiao.Huang, HouZhi. Zheng*, LiGuo.Wang, Lin.Chen and JianHua.Zhao

State Key Laboratory for Superlattices and Microstructures,
Institute of Semiconductors, Chinese Academy of Sciences,
P.O. Box 912, Beijing 100083, China

The expected features of diluted magnetic semiconductors still remain in controversial issue, concerning whether or not *s, p-d* (*f*) exchange interactions indeed modify the host semiconductor band structure. To solve this doubt, a new scheme for measuring magneto-optical (MO) effect is developed, called magnetic photoresistance circular dichroism (PR-MCD), which detects the differential photoresistance of materials between two circularly polarized excitations. That allows us to detect the MO effect induced only by interband transitions, and provide unambiguous evidence that the host semiconductor band structure is indeed modified by the strong exchange interactions. Our PR-MCD spectra also disclose intrigue features which may come from strong coupling correlation effect at very high manganese concentration limit.

Spintronic technology requires materials that allow control of both the charge and the spin degrees of freedom of carriers. Diluted magnetic semiconductors (DMS) have been considered to meet such a requirement. Extensive study on the prototype ferromagnetic semiconductor (Ga,Mn)As over the past decade has revealed a variety of unique features. However, the basic physics of this material, including the mechanism behind the ferromagnetism and the band structure modification, remains not yet fully understood[1, 2]. The most controversial debate is related to whether the impurity band (IB) is merged with the valence band (VB model) or is decoupled from it (IB model)[3-9]. For the IB model, the Zener double-exchange-type mechanism[10] is applicable, and makes it questionable if the valence band of (Ga,Mn)As could still be modified substantially. As a result, it is very desirable to have an experimental method that can only probe the band modification, induced by *p-d* exchange, without the influence from any other processes. Magneto-optical (MO) effects are widely used to characterize the mechanisms related to ferromagnetism[5]. However, the observed MO effects may stem from various physical origins. Of them, optical transitions, including interband transitions, defect-to-band transitions, intra-valence-band transitions, transitions in various defect complexes and intra-magnetic ion transitions, are the main causes for MO effects[11]. That makes the interpretation of the magnetic circular dichroism (MCD) spectra not as straightforward as expected [5,12].

We have developed a new method for measuring MO effect, called magnetic photoresistance circular dichroism (PR-MCD), which detects the differential photoresistance between right ($\sigma^+$)- and left ($\sigma^-$)-circularly polarized excitations.

That allows us to detect MCD effect caused only by interband transitions, and provide unambiguous evidence that the host semiconductor band is actually modified by the strong exchange interactions. Moreover, we show the influence of strong coupling correlation on PR-MCD.

PR-MCD is related to material susceptibilities in a manner as derived in the following. The photoresistance, induced by $\sigma^+$- and $\sigma^-$ excitations, is given by

$$\Delta R^\pm = \{e(\mu_n + \mu_p)\frac{W}{L}[\tau \eta I_0(1-|r|^2)(1-e^{-\alpha^\pm d})]\}^{-1} \quad (1)$$

, where $W$, $L$, $d$ are the width, length and thickness of (Ga,Mn)As epilayer, $\mu_n$, $\mu_p$ the mobilities of electrons and holes, $\tau$ the lifetime, $I_0$ the intensity of incident light at the surface, $|r|^2 \cong |r^+|^2 \cong |r^-|^2$ the surface reflectivity, $\eta$ the quantum efficiency and $\alpha^\pm$ the absorption efficiencies. PR-MCD is defined as their difference $\delta R = \Delta R^+ - \Delta R^-$, and approximately expressed in a weakly absorbing limit as

$$\delta R \cong -(\Delta\alpha/\alpha)/\{e(\mu_n + \mu_p)\frac{Wd}{L}[\alpha\tau\eta I_0(1-|r|^2)]\} \quad (2)$$

, where $\alpha = (\alpha^+ + \alpha^-)/2$ and $\Delta\alpha = \alpha^+ - \alpha^-$. Obviously, PR-MCD signal is in proportion to the relative difference $\Delta\alpha/\alpha$ between $\alpha^+$ and $\alpha^-$. By solving the eigenmodes of EM waves, propagating in an electrically gyrotropic and weakly absorbing medium, reference 13 derived at

$$\alpha^\pm = \frac{n\omega}{c}[(4\pi/\varepsilon)(\chi_e^{"xx} \mp \chi_e^{'xy})] \quad (3)$$

where $n$, $\varepsilon$ are the refractive index and dielectric constant of the medium, respectively, $c$ the speed of light, and $\omega$ the frequency of EM wave. Eq. (3) relates $\alpha^\pm$ to the elements of susceptibility tensor in Faraday configuration with $\chi_e^{"xx}$ being the imaginary part of diagonal element and $\chi_e^{'xy}$ the real part of off-diagonal element.

Finally, one derives

$$\Delta\alpha = \alpha^+ - \alpha^- = -\frac{8\pi n\omega}{\varepsilon c}\chi_e^{'xy} \qquad (4)$$

Therefore, PR-MCD is determined by the real part $\chi_e^{'xy}$ of the off-diagonal element of tensor $\overline{\overline{\chi}}$.

As a comparison, let us recall the corresponding expression of MCD in reflectance. Conventionally, MCD is defined as $(90/\pi)\left(|r^+|^2 - |r^-|^2\right)/\left(|r^+|^2 + |r^-|^2\right)$. Following reference 14, after some algebra, one finds that the MCD in reflectance depends on the imaginary part of the off-diagonal element $\varepsilon_e^{"xy} = \chi_e^{"xy}$ in the form

$$MCD \cong -(90/\pi)[2\varepsilon_e^{"xy}/n(n^2-1)] \qquad (5)$$

Here, $n = (n^+ + n^-)/2$, $n^\pm$ and $\kappa^\pm$ are the real parts and imaginary parts of the complex refractive indexes $N^\pm$ for $\sigma^+$- and $\sigma^-$- excitations.

Two kinds of (Ga, Mn)As samples were used in the present study. Sample A was grown by low-temperature molecular-beam epitaxy (LT-MBE), and consists of a 100nm-thick $Ga_{0.95}Mn_{0.05}As$ epilayer grown on the top of a semi-insulating $In_{0.19}Ga_{0.81}As$/GaAs composite substrate. Sample A was then annealed at 250 $^o$C in air for one hour, showing a Cure temperature $T_C$ of 110 K and a easy axis normal to the surface. Sample B consists of 70nm-thick $Ga_{1-x}Mn_xAs$ epilayer on a semi-insulating GaAs substrate and was post-annealed in air for 16 hours. The nominal mole concentration of Mn was $x = 20\%$. The easy axis of sample B lies in the plane.

For reflectance MCD measurements, the experimental setup used is conventional one, and described in supplemental. To measure PR-MCD and PR, two ohmic contacts were made at the two ends of the samples, which were fed by a constant current of 200 $\mu$A. The photo-voltage response was detected at the chopping frequency for PR signal or at 50 kHz for PR-MCD.

We first check if the magneto-crystalline anisotropy affects PR-MCD in the same way as it does in conventional MCD and anomalous Hall resistance (AHR). The hysteresis loops obtained from PR-MCD, MCD and AHR measurements have been plotted in Fig. 1 after normalizing by their respective saturated values. All of them were measured at 4.2 K and $\lambda$ = 750 nm (for PR-MCD, MCD). Because the polarity of MCD is opposite to both PR-MCD and AHR, the measured MCD is mirrored to allow easy comparison. Figure 1 shows that the three hysteresis loops are exactly the same with a same coercive field. This fact convinces us that the change in magnetization controls the band modification in the same way as it does in conventional MO effects and static magneto-transport.

To demonstrate the uniqueness of PR-MCD, PR-MCD, MCD and PR spectra were measured for sample A under a saturated magnetic field of 0.08 Tesla at different temperatures as shown in Fig. 2. The same measurements were also performed under a magnetic field of 0.7 Tesla for sample B, and are plotted in Fig. 3. As seen in Fig. 2 and 3, when the direction of the applied magnetic field is reversed, the spectra of PR-MCD and MCD change their sign. Their symmetry about the zero line indicates that the data are of high quality. The pronounced peaks at 825 nm (close to GaAs gap at 4.2 K) in both PR-MCD and PR spectra come from the photo-excitation in the substrate (for details, see supplemental).

It is clear from Fig. 2 that the overall aspect of PR-MCD is substantially different from that of MCD. While MCD signals persist far beyond $\lambda$ = 900 nm, there is a long wavelength cutoff in both the PR-MCD and PR spectra. Moreover, by referring supplementary, one finds that beyond the cutoff (~840nm), the hysteresis loop of PC-MCD of sample A at 4.2 K indeed collapses to a horizontal line. In contrast, the corresponding hysteresis loop of MCD remains clearly seen at $\lambda$ = 880 nm, 920 nm. The sample B also shows the same behaviors in Fig. 3. PR-MCD hysteresis loop collapses also at the wavelength longer than 840 nm (see supplemental).

In order to understand these features, we shall briefly discuss the possible mechanisms contributed to PR-MCD. The optical transitions within various defect complexes and intra-magnetic-ion transition do not create mobile carriers, and thus have no contribution to either PR-MCD or PR. As for the impurity-to-band transitions, one knows that, in addition to conventional defects (like arsenic antisites, arsenic interstitials, and their complexes), a substantial portion of the Mn impurities in LT-(Ga, Mn)As were found in interstitial positions[15]. All of these defects have high-lying levels in the gap, and act as donors, compensating for the Mn substitutive impurities, and decreasing hole concentration. The transitions from valence band to these defect levels would create mobile holes in the valence band and contribute both PR and PR-MCD. However, the response from these transitions is not found in either Fig. 2a,c or Fig. 3a,c.

The last but not least question is where the Fermi level ($\varepsilon_F$) locates. It dwells in the valence band merged by the manganese impurity band or is in the Mn impurity band separated from the valence band? This has been the focus of current controversy. In the framework of the IB model, optical transition from the VB to the Mn impurity band (IB) was thought to persist in a wide spectrum region because of the relaxation of the optical selection rule [5]. The mobile holes created in the valence band would simultaneously enhance PR and PR-MCD. However, neither PR nor PR-MCD measurements show a discernible trace of such a contribution in the spectrum range considered here, as seen in Fig. 2a,c and Fig. 3a,c.

When VB model is applied, the optical excitation of holes from the energy region above $\varepsilon_F$ to the deep of the valence band does not change the total number of holes. The mobility of the excited hole might be slightly different from that above $\varepsilon_F$ because of enhanced kinetic energy and changed effective mass. As a result, the contribution of the intra-valence band transitions to both PR and PR-MCD might not be detectable. In fact, the present PR-MCD results do not provide the evidence either for the optical excitation from the valence band to the Mn impurity band or for intra-valence band excitation. We can conclude that the PR-MCD measured here only arises from MO effects due to the interband optical transitions, and reflects purely the band modification effect.

PR-MCD also exhibits significant difference from conventional MCD in the spectrum range shorter than the absorption edge. PR-MCD signals never change their sign in a wavelength scan, while MCD signals always reverse sign as seen from both Fig. 2a,b and Fig. 3a,b. Such a contrast between PR-MCD and MCD is closely related to their different dependencies on the real $\chi_e^{'xy}$ and imaginary $\chi_e^{"xy}$ parts of the off-diagonal element of the susceptibility tensor. When MO effect is caused by the resonant transition of $\omega_0$ between two local levels, according to Ref. 14, it follows that $\chi_e^{'xy}(\omega)$ is symmetric and $\chi_e^{"xy}(\omega)$ is anti-symmetric about the resonant frequency $\omega_0$. The latter implies a sign reversal upon crossing $\omega_0$. When MO effect is induced by the band-to-band transition, $\chi_e^{'xy}(\omega)$ and $\chi_e^{"xy}(\omega)$ may still retain such a feature to some extent. That explains the distinct difference between PR-MCD and MCD spectra in a range above absorption edge. The other noticeable difference is that PR-MCD signals are suppressed much faster than MCD signals by increasing temperature. In particular, as seen in Fig. 3a-c, both PR-MCD and PC signals tend to disappear as the temperature rises to 75 K, while MCD signal remains almost unchanged. This indicates that the MCD signal above the absorption edge may be mixed with contributions other than from the interband transition.

There is another feature worthy of discussion. As seen in Fig. 2a,c and Fig. 3a,c, both PR-MCD and PR signals resume somewhat as the wavelength becomes shorter than 725 nm (1.71 eV) in sample A and 680 nm (1.82 eV) in sample B at elevated temperatures, respectively. Such bumps on the short wavelength side may reasonably be ascribed to the transition from the spilt-off valence band to the conduction band. Their energies are close to the values adopted in Refs. 5 and 16. Previously, it was thought that the sign of MCD signal at $E_\Gamma(T)$ was opposite to that at $E_\Gamma(T)+\Delta$ [5], because the split heavy and light hole bands are in a level order of hh (-3/2), lh (-1/2), lh (1/2) and hh (3/2), making the $\sigma^+$ transition energies smaller than the $\sigma^-$ transition energies. The situation becomes opposite for the transitions from the split-off hole band to the conduction band. For PR-MCD, both Fig. 2a and Fig. 3a show that its sign at $E_\Gamma(T)+\Delta$ is the same as that at $E_\Gamma(T)$. This looks somehow by contraries. However, the $\sigma^+$ absorptions of heavy and light holes may take place earlier or later than the corresponding $\sigma^-$ absorptions with increasing the hole-filling in the split heavy and light hole bands, as long as $\varepsilon_F$ is inside the valence band and $k_\parallel$ is conserved[16], making their sign of PR-MCD same as that from split-off hole band at high hole concentration.

From now on, we want to disclose a new feature of PR-MCD, not found by previous methods. In Fig. 4a, both remnant magnetization $M_r$ in [001](measured by SQUID) and resistance were measured as a function of temperature in sample A. As the temperature decreases continuously from $T_C$ of 110 K, the resistance keeps decreasing monotonically, indicating an entrance of metallic phase. By checking supplementary, one finds that the PR-MCD persists to a temperature (~90K) rather close to $T_C=110K$. Now we turn to examine the behaviors of sample B. The temperature dependence of $M_r$ was measured under zero field, after sample B was cooled down under a perpendicular magnetic field of 1 Tesla. We find that this sample displays a reentrant insulating phase as the temperature decreases blow ~80 K, as shown in Fig.4b. Such a reentrant insulating phase has been observed in $Ga_{1-x}Mn_xAs$ epilayers of higher $x$ values, and attributed to the localization of holes, arising from a modified screening of the Coulomb potential due to enhanced interaction and correlation [2,3,17,18]. However, the measured PR-MCD signal in Fig.

3 persists in such a reentrant insulating phase. Intriguingly, it disappears as long as the temperature rises above 75 K in spite of a $T_c$ as high as 150 K, while its magnetization collapses completely as well (see supplementary). Therefore, it appears that the band modification by *p-d* exchange does not seize up in the reentrant insulate phase. The strong coupling correlation, which is neglected in mean field theory, may possibly account for our observation [17]. Although we do not have the exact values for the Fermi energy $\varepsilon_F$ and $n/N$, following equations 6, 7 and figure 1 of reference 17 and putting the numbers of $\varepsilon_F = 0.1 eV$, $S = 5/2$ and $n/N = 0.1$ in, we get critical temperatures of $T_C^{RKKY} = 0.53 T_C^{MF}$ in RKKY weak coupling regime and $T_C^{sc} = 108 K$ in the strong coupling regime. As speculated by the theory, the strong coupling correlation completely polarizes the band-electron system, making PR-MCD observable in our experiments, while RKKY weak coupling correlation has only a weak perturbation on the band structure. Therefore, the strong coupling regime is more responsible for our case. The reason why PR-MCD of sample B disappears early at a temperature about half Curie temperature may stem from the following fact. Because PR-MCD is measured in a large piece of bulk material, near the reentrant insulate-metal transition, its signal may be more easily averaged out due to the presence of long wavelength magnetic fluctuations than it does in magnetization measurements. The simultaneous suppression of PR response is unexpected, and may possibly be related to the spatial fluctuation of density of states (DOS) and the opening of correlation gap at the Fermi level, as revealed by recent scanning tunnelling microscopy experiments [19]. That needs to be clarified in the future.

Acknowledgements: This work was partly supported by The National Basic Research Program of China under Grant 2011CB932901 and the National Natural Science Foundation of China under Grants 60836002.

Figure Captions

**Figure 1 (color online) Normalized PR-MCD, MCD and AHR of sample A.**
Normalized PR-MCD, MCD and AHR at wavelength of 750 nm are measured at 4.2 K for sample A in a magnetic scan between ±0.08 Tesla.

**Figure 2 (color online) Spectra of PR-MCD, MCD and PR of sample A.** Spectra of PR-MCD, MCD and PR, normalized by the respective magnitudes at 4.2 K, are measured for sample A under a saturated magnetic field of 0.08 Tesla at different temperatures.

**Figure 3 (color online) Spectra of PR-MCD, MCD and PR of sample B.** Spectra of PR-MCD, MCD and PR, normalized by the respective magnitudes at 4.2 K, are measured at different temperatures for sample B under a magnetic field of 0.7 Tesla.

**Figure 4** (color online) **Temperature dependence of remnant magnetization and resistance of samples A and B.**
**a,** Sample A was cooled down under zero field, then both remnant magnetization and resistance of sample A was measured as a function of temperature. **b,** Sample B was cooled down under a perpendicular field of 1 Tesla first, then both remnant magnetization and resistance of sample B was measured as a function of temperature in the absence of magnetic field.

Fig. 1

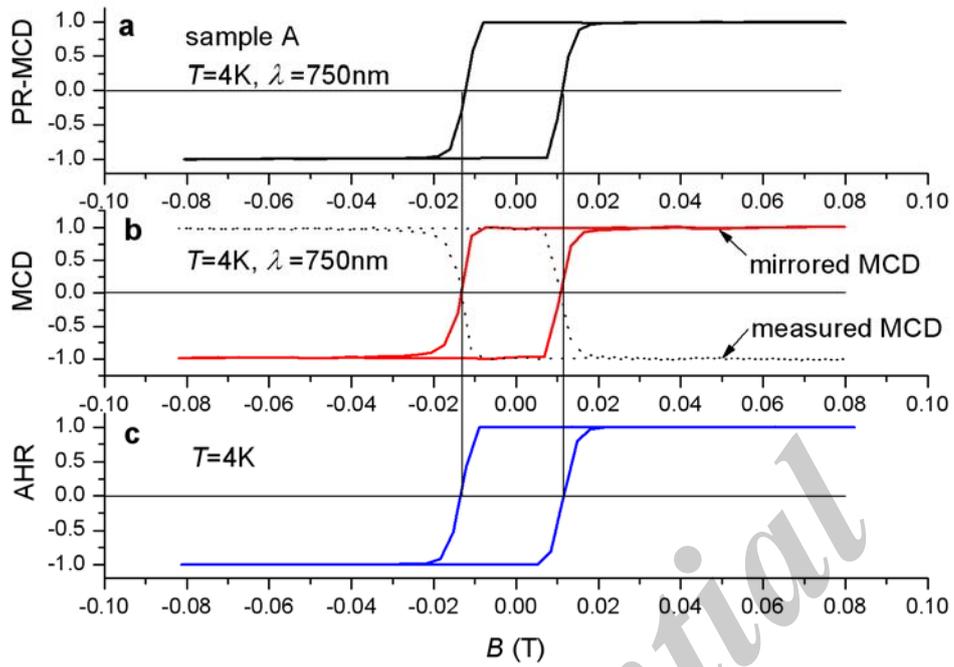

**Fig. 2**

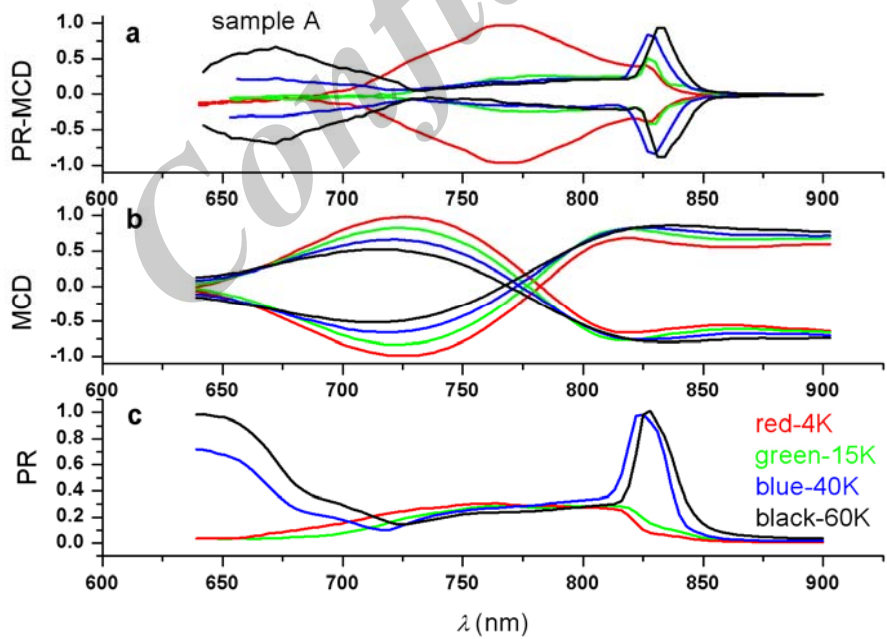

Fig. 3

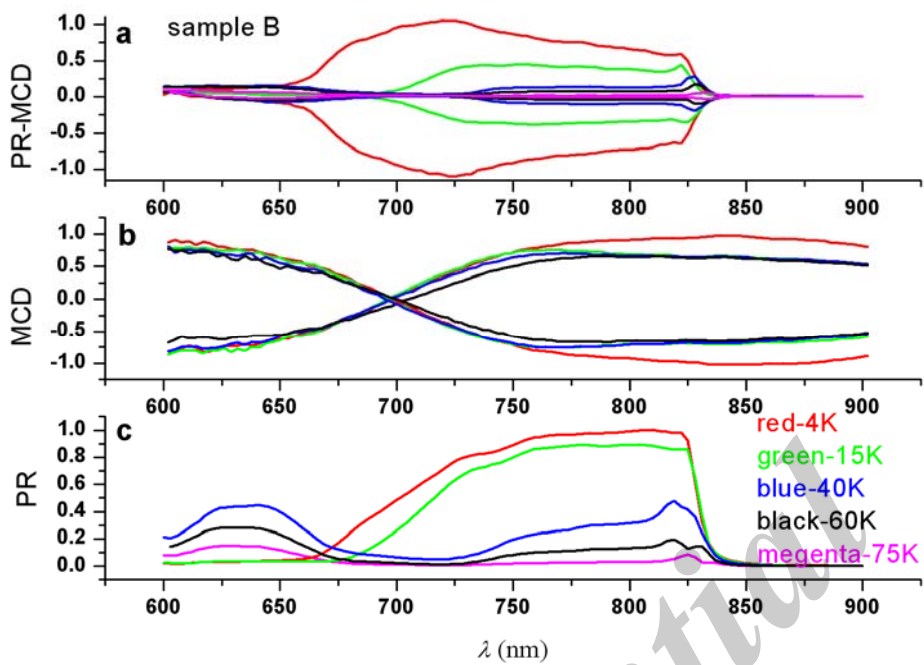

Fig. 4

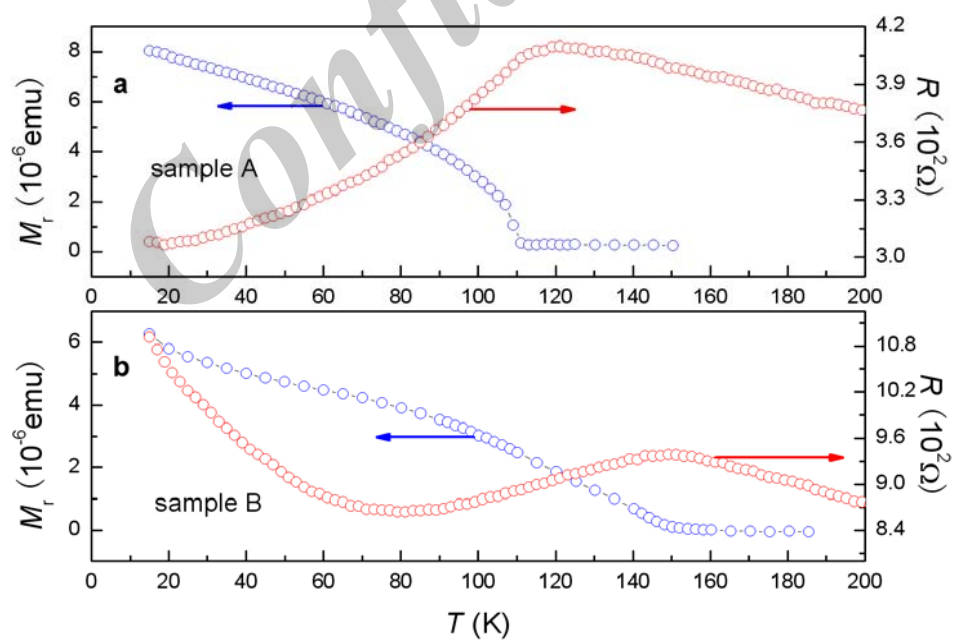